# Quantum Chemical Analysis of the Excited State Dynamics of Hydrated Electrons

By
P.O.J. Scherer and Sighart F. Fischer
Physik Department T38, Technische Universität
München, Germany

**Abstract:**

Quantum calculations are performed for an anion water cluster representing the first hydration shell of the solvated electron in solution. The absorption spectra from the ground state, the instant excited states and the relaxed excited states are calculated including CI-SD interactions. Analytic expressions for the nonadiabatic relaxation are presented. It is shown that the 50fs dynamics recently observed after s→p excitation is best accounted for if it is identified with the internal conversion, preceded by an adiabatic relaxation within the excited p state. In addition, transient absorptions found in the infrared are qualitatively reproduced by these calculations .

**Introduction**

While the absorption spectrum of the hydrated electron can be well described by different theoretical approaches such as MD simulations including one electron pseudo potential approximations /1-5/ or using density functional theory /6/ there are still unsolved questions concerning the interpretation of the initial 50fs kinetics after an s-p excitation, either as adiabatic relaxation (AR) /1-5,7,8/ or as internal conversion (IC) to the ground state /9/. A very good time resolution as well as a good frequency resolution is needed from the experimental side, to resolve this problem. Recent pump- probe studies by Pshernichnikov et. al /9/ provide a time resolution with 5fs pulses within a frequency range of 10 000 $cm^{-1}$ to 17 000 $cm^{-1}$. They interpret the observed 50fs relaxation as the IC-process to the ground state. The analysis of the time dependent spectra includes excited state absorption, stimulated emission and a decay to an unrelaxed ground state with its p-type excitation. They do not explicitly include any excited state relaxation, which indeed does not show up as a red shifted stimulated emission within the frequency window of their excitation and probing pulses. For deuterated water they find a lengthening of the IC - lifetime up to 70fs. The adiabatic relaxation model, on the other hand, seems to be in line with experiments by Thaller et.al./7/, which show spectral features in the range of 4000$cm^{-1}$, consistent with the transient relaxing spectrum from an excited p state.

For a microscopic analysis, a good knowledge of the excited state absorption spectrum is needed as function of the relaxation coordinates. To approach this problem we restricted ourselves to the first solvation shell in symmetry conserving configurations and relaxed the hydrogen atoms for fixed O-atom positions to achieve the ground state geometry. The different excited states are calculated for this geometry. After excitation of one of the p-states, again relaxation of the H-atoms is invoked. We argue that only the H-atoms can respond substantially on the 50fs time scale. The transient spectra for the relaxed configuration are then also evaluated and it is assumed that the internal conversion proceeds from this configuration. With these assumptions, time resolved transient spectra are predicted consistent with the measurements of /9/. In addition, the transient spectra observed in /7/ are nicely reproduced and reinterpreted within the same relaxation scheme, which identifies the 50fs relaxation time as an internal conversion process, preceded by an adiabatic relaxation, in a time range of 20fs. Finally, we present a model to evaluate the rate for the internal conversion. It involves a nonadiabatic coupling resulting from promoting modes which modulate the molecular dipoles and couple via Coulombic interaction to the electronic transition dipole.

**Quantumchemical methods:**

The calculations were done with the GAMESS program package /11/. We used the TZV (triple zeta valence) basis with additional polarization and diffuse basis functions. The excited state spectra were obtained by CI-SD calculations including 15 occupied and 15 virtual orbitals.

Normal modes were calculated for the ground state with the O-atoms fixed.

**Results:**

**a) Transient Spectra**

The ground state geometry was optimized on the HF level. Inclusion of MP2 did not change the structure significantly. The oxygen atoms were kept fixed in a tetrahedral arrangement. After optimizing all the hydrogens the resulting structure showed approximately $S_6$ symmetry which is a subgroup of the full tetrahedral symmetry. (Fig 1) . Table 1 gives the structure parameters for the solvated electron in the relaxed ground state and for relaxed excited states. Due to the structure of the water molecule this is the highest symmetry possible for a $(H_2O)_6$ cluster with the oxygens in a tetrahedral arrangement and the two hydrogens not equivalent. The irreducible representations of $S_6$ are $A_g, A_u, E_g$ and $E_u$. We discuss our results within this symmetry. This way we reduce the dimension of the degrees of freedom for the quantum calculation and can classify the stable minima. Especially for the spherical harmonics we note, that functions $Y_{l,m}$ and $Y_{l,-m}$ remain degenerate (they correspond to the E representations of $S_6$). Therefore p-orbitals are split into $p_z$ and the pair $p_x, p_y$ whereas d-orbitals are split into $d_{z^2}$ and two pairs $d_{x^2-y^2}, d_{xy}$ and $d_{xz}, d_{yz}$.

Fig 2 shows the orbitals relevant for the ground and the lowest excited states. They are denoted in analogy to the H-atom as s,p,d – orbitals according to their symmetry. For the transitions configuration interaction is included. The ground state absorption spectrum from its relaxed configuration (s relaxed) is shown in Fig 3a. Only the transitions s→$p_z$ and s→ ($p_x$, $p_y$) denoted as $p_A$ and $p_E$ , to classify them according to their symmetry, carry substantial intensity . Compared to experiments in the bulk they should be shifted downwards in energy by about 1500cm$^{-1}$. In the following we shall refer also to the energetic location of transitions relative to the central P$_A$- peak which is located in the calculation at 15000cm$^{-1}$ . The weak transitions 14500 cm$^{-1}$ above the central peak are of s→f-type. They are imbedded in the continuum and have been excited by single photon ionization experiments /12/. As can be seen the p$_A$ excited state should represent the long wavelength side of the experimental p-band, since it lies 1 300 cm$^{-1}$ below the degenerate p$_E$ states. Fig 3b and Fig 3c show the transient spectra starting from p$_A$ and p$_E$, respectively for the same s-relaxed configuration (instant excited). We like to point specially at the low energy transition to a 2s-type state located at 3850 cm$^{-1}$ (Fig 3b). It correlates nicely to absorbances observed in this frequency regime by Thaller et.al. /7/. The transient spectrum starting from the unrelaxed p$_E$ state (Fig 3c) shows less intensity for the lowest 2s type transition (Fig.3b) but more for the lowest d$_E$ type transition located at 6500 cm$^{-1}$.

Next we searched for the nuclear configuration that minimizes the lowest excited state, denoted p$_A$ relaxed. This relaxation induces strong shifts in the s→p transitions (Fig 4a) and in the transient spectra (Fig 4b). We find a lowering of the p$_A$→s stimulated emission energy by 5800 cm$^{-1}$ (Fig 4a) and an increase in the transient absorption energy p$_A$→2s, ,d$_E$ d$_A$ by 2300 cm$^{-1}$ up to 6000 cm$^{-1}$ respectively (Fig 4b) . The splitting of the p$_E$ and p$_A$ states increases up to 5200 cm$^{-1}$. Relaxation of the p$_E$ state causes a similar downshift of that state but not a comparable upshift of the transient spectrum.

**b) Simulation of the dynamics**

In our simulation of the time dependent spectra we used a band shape function and empirical relaxation times as proposed in /9/ for the combined s→$p_A$ and s→$p_E$ transitions, in order to get a good basis for the comparison with their simulation. The optical excitation from the ground state can reach either the $p_A$ state only (long wavelength excitation /7/ Fig 5a )or the $p_A$ and $p_E$ states weighted by the factor two (broadband excitation /9/ Fig 5b). The time evolution of the spectra is calculated for the following reaction model

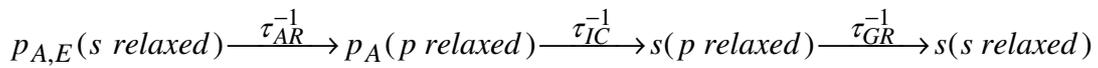

$$p_{A,E}(s\ relaxed) \xrightarrow{\tau_{AR}^{-1}} p_A(p\ relaxed) \xrightarrow{\tau_{IC}^{-1}} s(p\ relaxed) \xrightarrow{\tau_{GR}^{-1}} s(s\ relaxed)$$

(1)

We include an adiabatic relaxation process with the relaxation time $\tau_{AR}$ =20fs to the optimal configuration for the lowest excited state which is of $p_A$ symmetry ($p_A$ relaxed) for both types of excitation. This time $\tau_{AR}$ corresponds to the initial Gaussian component of the dielectric relaxation function /3,14/. Within our model (see also section c) it relates to the early time contribution of the low frequency librational mode of 364cm$^{-1}$, which is responsible for the main geometry change (see Table 3). It will be strongly damped in an extended system with flexible oxygen positions. The second relaxation describes the internal conversion to the electronic ground state with the time constant $\tau_{IC}$ = 50fs, which is followed by a slower ground state relaxation with $\tau_{GR}$=250fs to the initial configuration (s relaxed). Within our model (section c) we can only explain $\tau_{AR}$ and $\tau_{IC}$ since they are dominated by the motion of the hydrogen atoms, not the oxygens. The ground state relaxation $\tau_{GR}$ describes the much slower reconstruction of the s relaxed structure, which will experimentally be disrupted also with regard of the oxygen positions due to the large energy release.

The experimental spectra of /9/ show characteristic changes, due to relaxation and nonadiabatic decay, which are reproduced by our simulations (Fig 5b) where the $p_A$ and $p_E$ states are excited with a probability ratio of 1:2 . The best marker of the internal conversion is the build up of a peak with $\tau_{IC}$ = 50fs about 1500cm$^{-1}$ below the central peak. It results to large extent from the s→$p_E$ transition (Fig 4a) which relaxes together with the s→$p_A$ transition to the corresponding transitions of Fig 3a. The bleach at the center evolves on all three timescales. 15% disappear with $\tau_{AR}$ , due to the shift of the stimulated emission next followed by a comparable contribution on the $\tau_{IC}$ timescale. Finally, the ground state population has to relax with $\tau_{GR}$ to restore the initial absorption. On the time span of our simulation of 100 fs about 1/3 of the bleach has recovered. A signature of the adiabatic relaxation is also seen as a decrease of the transient absorption about 3000 cm$^{-1}$ above the central peak. It moves towards higher energies on the timescale of $\tau_{AR}$ .The fact that a red shifting stimulated emission from $p_A$ is not seen in the experiments /9/ can be explained by the rapid relaxation $\tau_{AR}$, which is not well time resolved experimentally and the magnitude of the red shift which shifts the corresponding stimulated emission out of the experimentally detected frequency range. The time dependence of the low energy $p_A$ →2s transition is well seen in the $p_A$ excited spectra of Thaller et.al. /7/ (Fig 5a,5c) even though it is not fully time resolved. The observed lifetime $\tau_1$ from /7/ is , in our view, consistent with the rapid adiabatic relaxation with $\tau_{AR}$ followed by the $\tau_{IC}$ = 50 fs-(70fs for $D_2O$) nonadiabatic decay to the ground state. The larger amplitude of the short time absorbance observed for the deuterated system could result from the longer $\tau_{IC}$ lifetime of the p-state for this system.

To distinguish between the adiabatic and the nonadiabatic relaxation models in our simulation, we eliminated the IC - process for the $p_A$ excited system (Fig 5c): As can be seen, the main marker for the internal conversion, the growing peak below the central peak, is absent within this scheme. The changes in the low energy regime relax with $\tau_{AR}$ which is below the time resolution of /7/ . So we conclude, the nonadiabatic relaxation model with the 50fs internal conversion preceded by a 20fs adiabatic relaxation provides a consistent picture also for the $p_A$ excited system from /7/ simulated in Fig 5a.

**c) Relaxation and internal conversion**

Our model allows a more detailed analysis of the dynamics following the excitation in terms of stable normal modes, not just instant normal modes /16/. There are six local modes for every water molecule keeping the oxygens at fixed positions. The delocalized vibrations can accordingly be characterized by their symmetry and their local character, that is O-H stretch, H-bend and rotations, which become librations. So we calculated one bending, two stretching and three libration modes in every symmetry class respectively. The frequencies of the $A_g$ and $E_g$ modes are shown in Table 2 together with the displacements $g_k$ of the equilibrium positions of the $A_g$ modes in the excited $p_A$ state relative to the ground state equilibria given in units of the zero point amplitude of the normal mode in the ground state. The values

exp($-g_k^2$) gives the Franck-Condon transition probabilities for the 0-1 excitations of these modes in the absorption spectrum. They explain the asymmetry of the largely homogeneously broadened bandshape. The most strongly coupling $A_g$-mode is the collective libration mode at 364cm$^{-1}$ which controls the $\tau_{AR}$ relaxation. The $A_u$ and $E_u$ modes are listed in Table 3 together with the mode specific coupling constants squared and the partial IC- rates $k_k$. In comparison to experiments we get frequency values of the OH stretching modes slightly too high except for the $A_g$ mode at 2438cm$^{-1}$. A comparison with results from cluster work is not meaningful, since the geometry of our cluster is not relaxed as isolated entity. Finally we want to determine $\tau_{IC}$ within our model. In principle, we can determine the nonadiabatic couplings for the p→s transitions numerically. This however, would require substantial computational effort and be less instructive than the following approximation. Let us consider a single particle approach for the solvated electron in a pseudo potential

$$V(r) = \sum_i \frac{-eq_i}{4\pi\varepsilon|\vec{r}-\vec{R}_i|} + \sum_{i<j} \frac{q_i q_j}{4\pi\varepsilon|\vec{R}_i-\vec{R}_j|} \qquad (2)$$

The $q_i$ are effective charges at the hydrogen or oxygen atoms at positions $\vec{R}_i$. We have taken the value of q = 0.27 for both hydrogen atoms. The dielectric constant, which accounts for optical polarization is $\varepsilon = 1.7\varepsilon_0$. We denote the Born-Oppenheimer wave functions as $\varphi_\alpha(r;\{Q_k\})$ and the eigenfunction for the nuclear motion of mode k by $\chi_{\alpha,kn}(Q_k)$, where α represents the electronic state s or p and kn stands for the quantum state n of the normal mode $Q_k$ with frequency $\omega_k$ in the ground state. A p→s transition can only be induced by modes of $A_u$ or $E_u$ symmetry. Since these do not change their equilibrium position we neglect the dependence of the wave-functions $\chi_{\alpha,kn}(Q_k)$ on the electronic state $\alpha$. Thus we get as nonadiabatic matrix element for one promoting mode k of $A_u$ or $E_u$ symmetry (neglecting at this point the Franck Condon reduction of the $A_g$ modes).

$$V_{p,k,0 \to s,k,1} = -\sum_i \frac{\hbar^2}{M_i} \int dQ_k \chi_{k0}(Q_k) \int d^3r \left( \varphi_p(r;Q_k) \nabla_{Ri} \varphi_s(r;Q_k) \right) \nabla_{Ri} \chi_{k1}(Q_k) \quad (3)$$

Taking for this mode only on shell energy changes $E_p(Q_k) - E_s(Q_k) = \hbar\omega_k$ we can transform the matrix element to a diabatic coupling term

$$V_{p,k,0 \to s,k,1} = \frac{1}{\hbar\omega_k} \sum_i \frac{\hbar^2}{M_i} \int d^3r \; \varphi_p(r) \frac{eq_i(\vec{R}_i - \vec{r})}{4\pi\varepsilon |\vec{R}_i - \vec{r}|^3} \varphi_s(r) \int dQ_k (\chi_{k0} \nabla_{Ri} \chi_{k1}) \quad (4)$$

The electronic wavefunctions are taken here for the p-relaxed configuration. The electronic integral can easily be integrated if we apply a multipole expansion with respect to r.

$$V_{p,k,0 \to s,k,1} = \sum_i \sqrt{\frac{\hbar}{2M_i\omega_k}} \frac{\vec{u}_{ki} q_i}{4\pi\varepsilon |\vec{R}_i|^3} \left( 3 \frac{\vec{R}_i}{|\vec{R}_i|} (\vec{\mu} \frac{\vec{R}_i}{|\vec{R}_i|}) - \vec{\mu} \right) \quad (5)$$

Here $\vec{u}_{ki}$ denotes the normal mode eigenvector which relates the gradient with respect to the nuclear coordinates and the normal mode elongation by

$$\vec{\nabla}_{Ri} = \sqrt{M_i} \sum_k \nabla_{Qk} \vec{u}_{ki} \quad (6)$$

and $\vec{\mu} = \int \varphi_s(\vec{r})(-e\vec{r})\varphi_p(\vec{r}) d^3r$ is the electronic transition dipole moment. If we expand further around the average positions of the water molecules $\vec{R}_m^0$ we have

$$V_{p,k,0 \to s,k,1} = \sum_{m=1..6} \frac{\vec{\mu}_{km}^{IR}}{4\pi\varepsilon(R^0)^3} \left( 3 \frac{\vec{R}_m^0}{R^0} (\vec{\mu} \frac{\vec{R}_m^0}{R^0}) - \vec{\mu} \right) \quad (7)$$

$\vec{\mu}_{km}^{IR}$ is the contribution of molecule m to the IR transition dipole

$$\vec{\mu}_k^{IR} = \sum_{m=1..6} \vec{\mu}_{km}^{IR} = \sum_i \sqrt{\frac{\hbar}{2M_i\omega_k}} q_i \vec{u}_{ki} \quad (8)$$

The $p_A \rightarrow s$ transition, which is polarized along the symmetry axis, couples only to $A_u$ modes and the perpendicular $p_E \rightarrow s$ transition only to the $E_u$ modes. Since the average distance $R^0$ is about 3Å, which is comparable to the electronic radius, this expansion is not well justified for the first shell. We approximate the electronic wavefunctions by Gaussian s and p functions in the $p_A$ relaxed configuration and determine the radius a=2.6Å so that it gives the same transition dipole as we calculated on the HF/CI level. The integrals from Eq.(4) can be evaluated analytically and we arrive at the following formula for the coupling matrix elements.

$$V_{p,k0 \rightarrow s,k,1} = \sum_i \sqrt{\frac{\hbar}{2M_i \omega_k}} \frac{q_i}{4\pi\varepsilon} (u_{si} \nabla_{R_i})(\mu \nabla_{R_i}) \frac{erf(|R_i|/a)}{|R_i|} \quad (9)$$

$$= \sum_i \sqrt{\frac{\hbar}{2M_i \omega_k}} \frac{q_i}{4\pi\varepsilon} \left\{ \frac{3(\vec{u}_{ki}\vec{R}_i)(\vec{\mu}\vec{R}_i) - R_i^2 \vec{u}_{ki}\vec{\mu}}{|R_i|^5} erf(|R_i|/a) + \right.$$

$$\left. + \frac{1}{\sqrt{\pi} R_i^2} \left( \frac{2\vec{u}_{ki}\vec{\mu}}{a} - 4\frac{(\vec{u}_{ki}\vec{R}_i)(\vec{\mu}\vec{R}_i)}{a^3} - 6\frac{(\vec{u}_{ki}\vec{R}_i)(\vec{\mu}\vec{R}_i)}{R_i^2 a} \right) e^{-R_i^2/a^2} \right\}$$

In the limit a<< R we can expand the error function and recover Eq.(7) as the leading term. Finally we apply a Gaussian approximation for the Franck-Condon weighted density of states and arrive at the following result with the sum going over $A_u$ or $E_u$ modes for the $p_A$ or $p_E$ states respectively

$$k = \sum_k k_k = \frac{2\pi}{\hbar} \sum_{k,s} \frac{|V_{p,k,0...s,k,1}|^2}{\sqrt{2\pi\sigma^2}} e^{-\frac{(\Delta E - \omega_k)^2}{2\sigma^2}} \quad (10)$$

Here $\Delta E$ stands for the energy difference of the p and the s-state in the p-relaxed configuration. We found the value $\Delta E$ = 9165 cm$^{-1}$, which is an upper limit, since we did not search with all combinations of displacements of the coupled $A_g$ modes for the minimum. $\sigma^2$ stands for the variance of the Franck Condon distribution squared.

In the low temperature limit $kT << \hbar\omega_{min} = 95 cm^{-1}$ it reads in a harmonic oscillator model with displaced equilibria $\sigma^2 = \sum g_k^2 (\hbar\omega_k)^2$. So we get with the $g_k$-values from Table 2 the result $\sigma^2 = 5 \times 10^6 cm^{-2}$.

We restricted our calculation to the zero temperature limit since the normal mode analysis for thermally activated low frequency libration modes would obscure the transition dipoles. They are strongly anharmonically coupled and turn into rotations at higher temperatures which provide a quite different temperature dependence as the normal modes.

The final rate expression from Eq. 10 with the coupling constants from Eq.(9) amounts to k = 10.4 (fs)$^{-1}$ corresponding to a lifetime of 96 fs if the system relaxes into the $p_A$ state as we assumed. The internal conversion from a pE relaxed state, induced by the $E_u$ modes would be even faster. Several corrections should be considered if we want to relate the results to measurements in the bulk. First, one has to incorporate long-range couplings to water molecules outside the first shell. In this case the dipole-dipole approximation can be used and the orientation of the molecules becomes random. This situation can be treated also analytically /15/ but it would extend the scope of this paper. Roughly speaking, we get a reduction of the lifetime by a factor of two which would bring us closely to the observed value of 50fs from /9/. While it is difficult to predict the absolute value of the rate more precisely than within a factor of two, we think that the deuteron effect gives a good test of the theory. As can be seen from the coupling matrix elements of Eq (9), we predict a lengthening of the lifetime by the factor of $\sqrt{2}$ due to deuteration.

A somewhat larger factor results, if the deuteron effect on the Franck Condon reduction factor is also included.

In summary, we would like to point out that the simulation of the transient spectra gives new support for the model, which identifies the 50 fs relaxation as the internal conversion process. Our coupling mechanism leads to a value of the rate, consistent to this assignment and it predicts a lengthening of the excited state lifetime for deuterated water by a factor of $\sqrt{2}$ in good agreement with observations /8,9/. These predictions are difficult to verify by MD calculations, which, according to our experience, cannot account properly for the quantum mechanical energy transfer coupling between the electronic and the vibronic excitations if for instance an algorithm as developed by Webster et al /15/ would be applied to the low temperature limit.

**Acknowledgements**

The authors like to thank A.Laubereau, A.Thaller, A.A.Zharikov and W.Dietz for fruitful discussions. The work has been supported by the DFG (SFB533).

Table 1: structure parameters for the equilibrium configurations of the s,$p_A$,$p_E$ states respectively as defined in Figure 1

|  | $r_1$ (Å) | $r_2$ (Å) | $\gamma$ | $d_1$ (Å) | $d_2$ (Å) | $\phi$ |
|---|---|---|---|---|---|---|
| s | 0.942 | 0.946 | 104.9° | 2.16 | 3.02 | 48.8° |
| $p_A$ | 0.947 | 0.936 | 106.6° | 2.23 | 3.12 | 93.8° |
| $p_E$ | 0.942 | 0.942 | 105.6° | 2.63 | 2.63 | 2.3° |

Table 2: frequencies of the $A_g$ and $E_g$ modes together with changes of the equilibria $g_k$ due to excitation $s \rightarrow p_A$ given in units of the zero point amplitudes. The $g_k$ values for the $E_g$ modes are zero

| $A_g$ $\omega_k$ (cm$^{-1}$) | 95 | 220 | 364 | 1013 | 2438 | 3997 |
|---|---|---|---|---|---|---|
| $g_k$ | 0.17 | 0.5 | 2.0 | 0.25 | 0.95 | 1.1 |
| $E_g$ $\omega_k$ (cm$^{-1}$) | 600(*) | 205 | 333 | 1653 | 3961 | 4029 |

Table 3: Frequencies of the Au and the Eu modes together with the corresponding IC-coupling constants squared and the partial rates

| $A_u$ $\omega_k$ (cm$^{-1}$) | 113(*) | 163 | 282 | 1612 | 3927 | 4036 |
|---|---|---|---|---|---|---|
| $V^2$ (eV$^2$) | 2.5x10$^{-2}$ | 3.9x10$^{-3}$ | 6.8x10$^{-3}$ | 7.4x10$^{-3}$ | 5.6x10$^{-3}$ | 6.8x10$^{-7}$ |
| K(ps$^{-1}$) | 2.4 | 0.003 | 0.59 | 4.1 | 3.3 | 0.005 |
| $E_u$ $\omega_k$ (cm$^{-1}$) | 129(*) | 274 | 323 | 1634 | 3914 | 4057 |
| $V^2$ (eV$^2$) | 1.0x10$^{-2}$ | 3.1x10$^{-3}$ | 5.4x10$^{-3}$ | 3.9x10$^{-3}$ | 3.0x10$^{-3}$ | 7.6x10$^{-4}$ |
| k(ps$^{-1}$) | 0.6 | 0.2 | 0.4 | 2.1 | 18.0 | 5.0 |

**Figure Captions**:

**Fig 1**

$S_6$-structure of the first shell water molecules in the relaxed ground state (s relaxed). The symmetry axis z points from the center of the electron through the corner of the dice with the oxygen atoms at the centers of the surfaces. The distances and angles shown for the selected water molecules refer to the s relaxed state.

**Fig 2**

Orbitals representing the ground state s and the lowest excitations of p and d-type
The figure was generated with MOLEKEL 4.3 (P. Flükiger, H.P. Lüthi, S. Portmann, J. Weber, Swiss Center for Scientific Computing, Manno ,Switzerland 2000-2002).

**Fig 3**

a) The ground state absorption spectrum . The transitions are denoted by the symmetry of the final state.

b) and c) Transient spectra starting from the $p_A$ and $p_E$ states respectively. The notation refers again to the final states.

All spectra refer to the s-relaxed configuration (instant transitions).

**Fig 4**

a) Ground state absorption for the configuration of the $p_A$ relaxed configuration.
b) Transient spectrum from $p_A$ in its $p_A$-relaxed configuration.
c) Transient spectrum from $p_E$ for its own $p_E$ -relaxed configuration.

**Fig 5**

Simulated time dependent transient spectra shown in time steps of 20fs starting with the dashed line.

a) Time evolution after excitation of the $p_A$ state only.
b) Time evolution after excitation of the $p_A$ state and the $p_E$ state with the probability ratio 1:2
c) Time evolution as in a) however without internal conversion ($\tau_{IC} = \infty$)

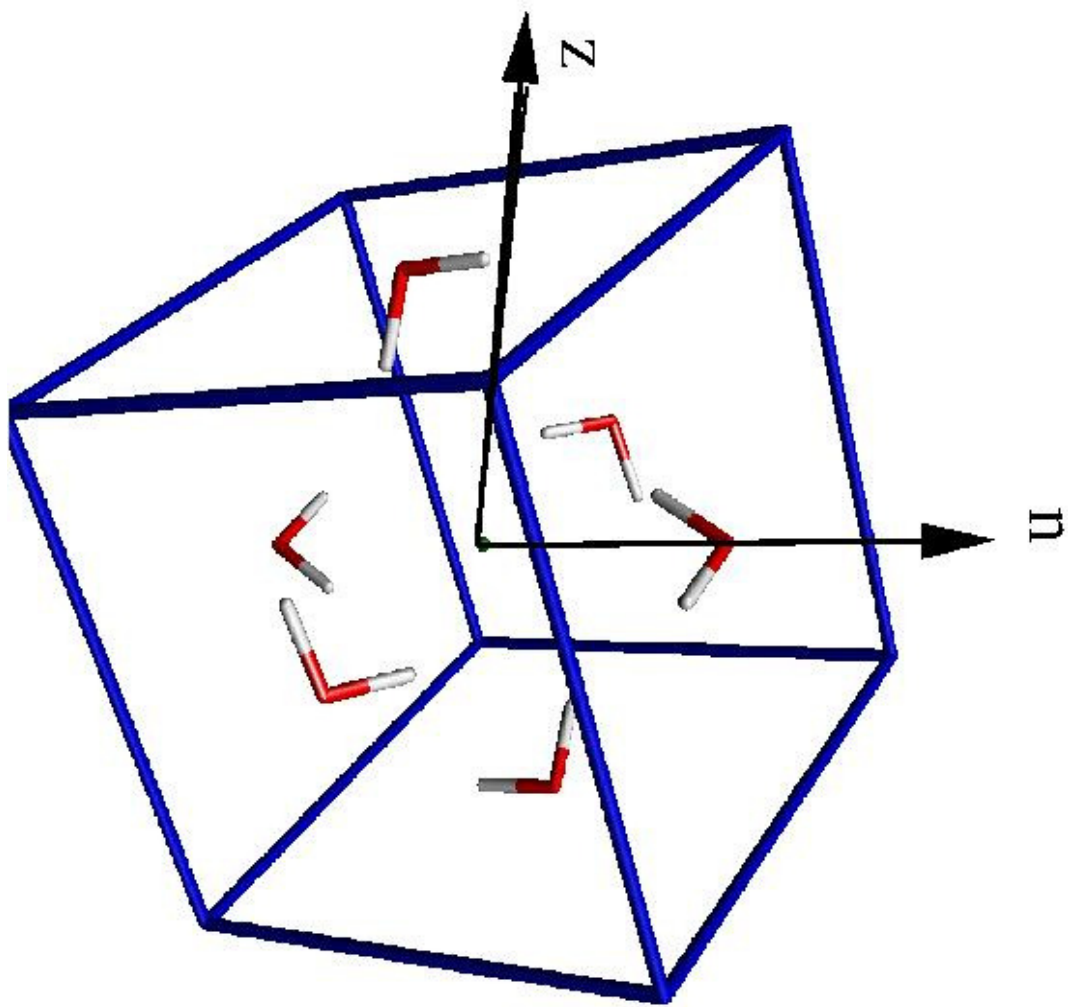

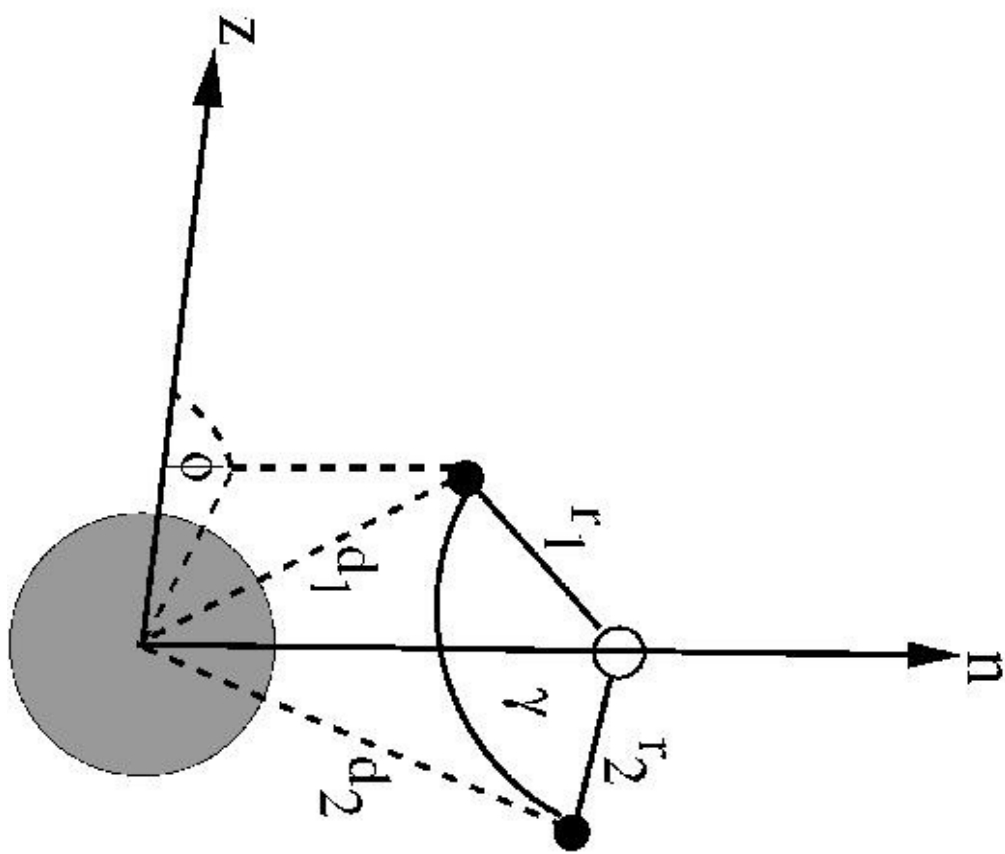

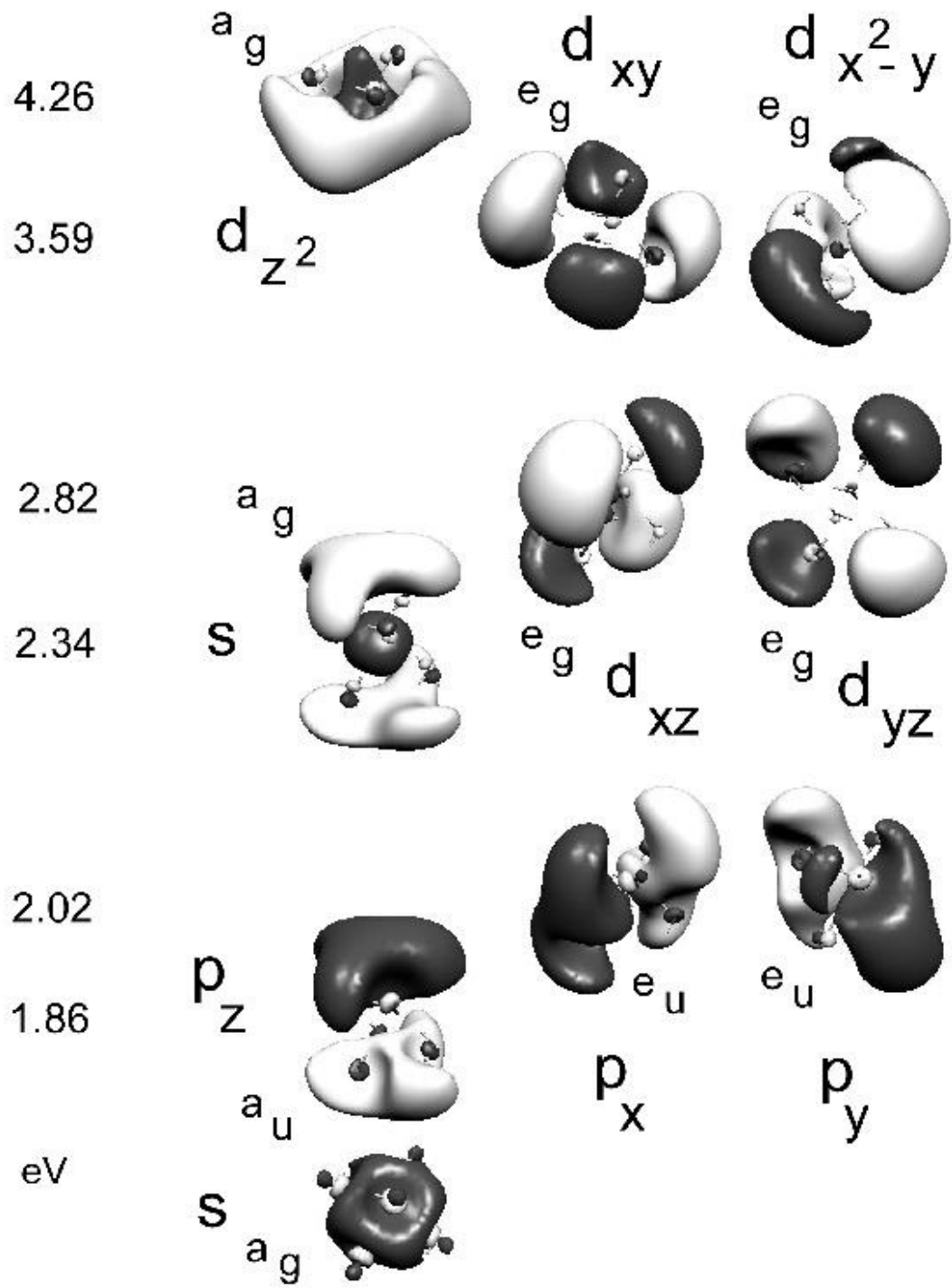

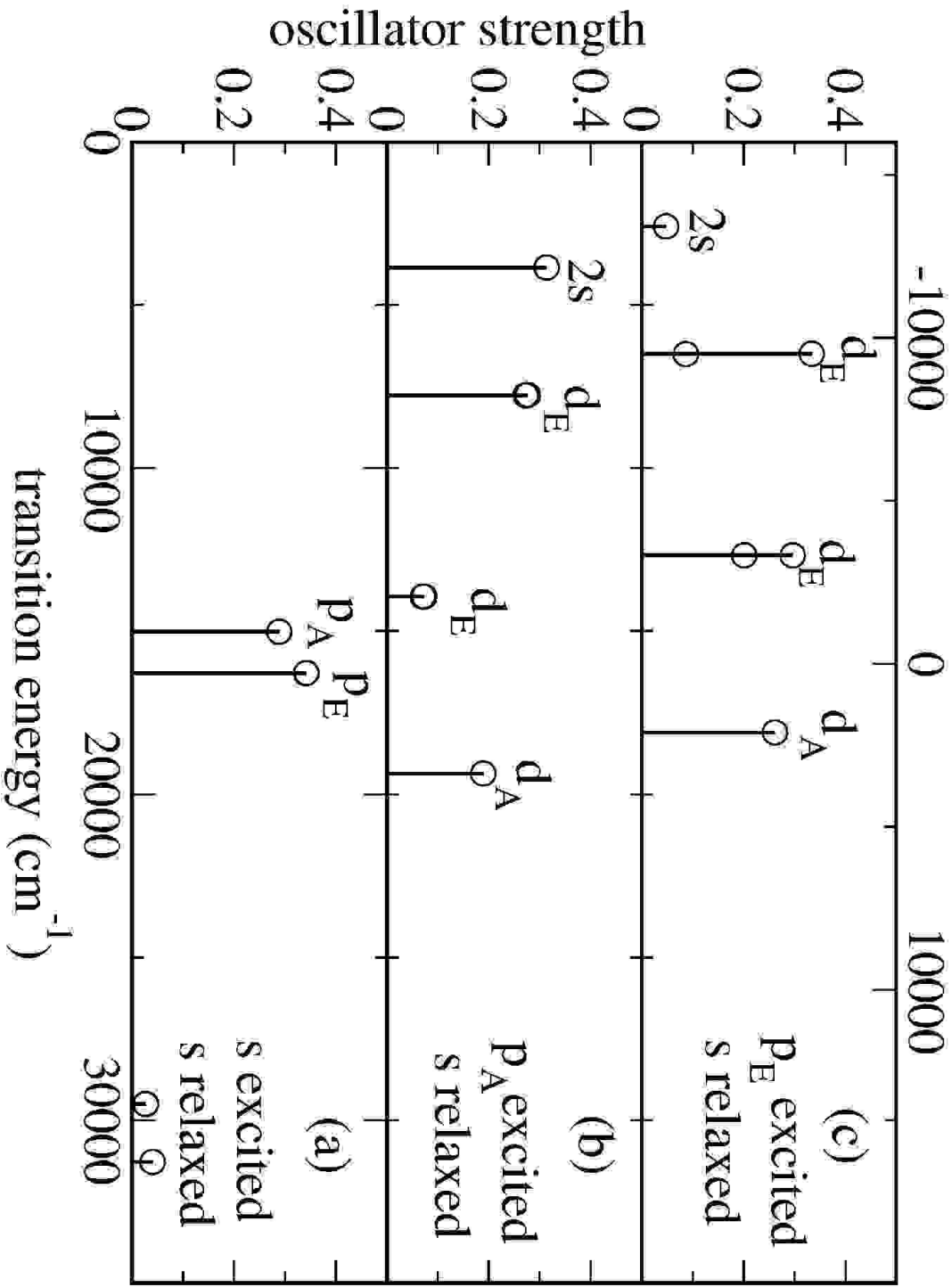

fig.3

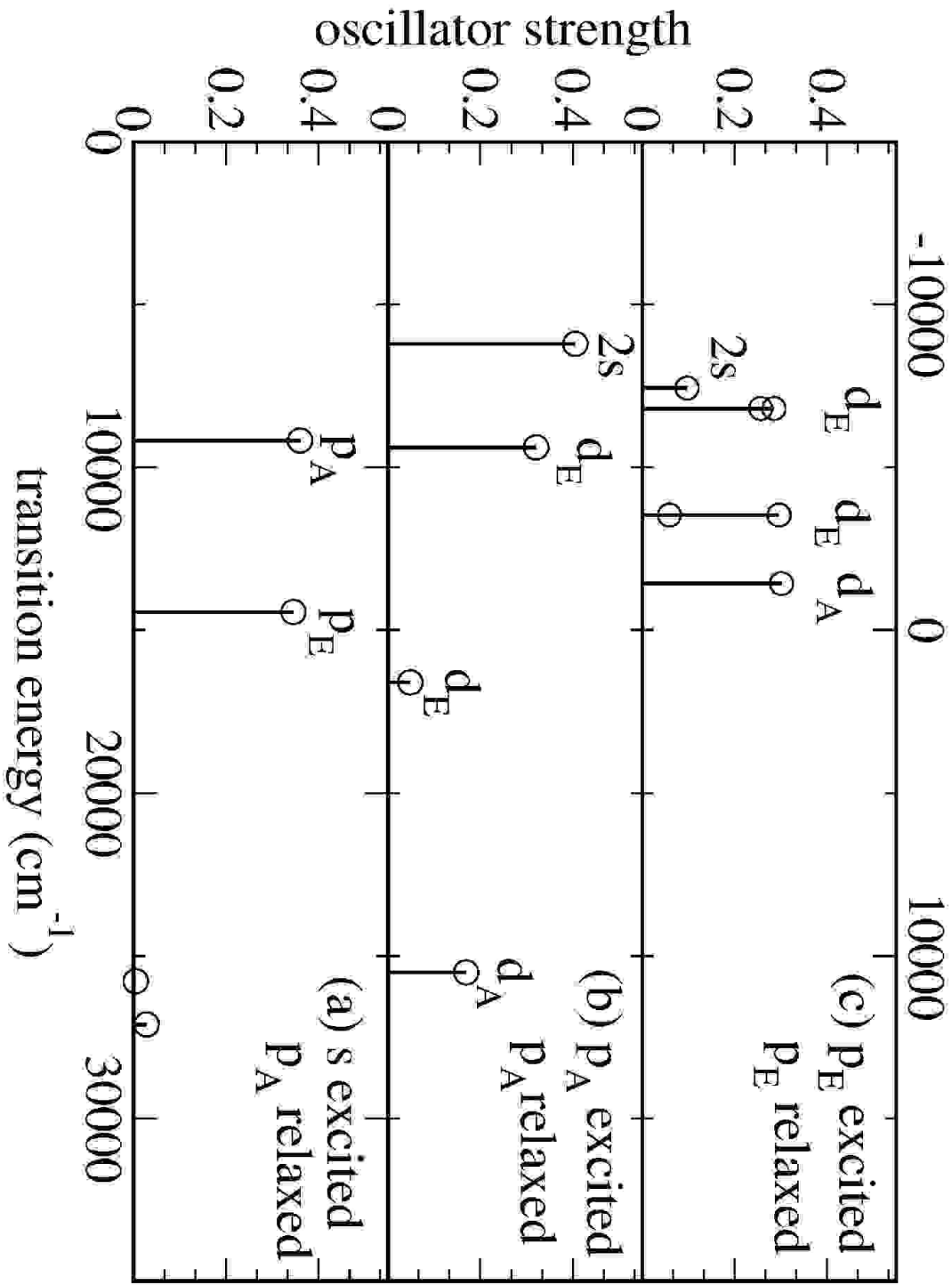

fig.4

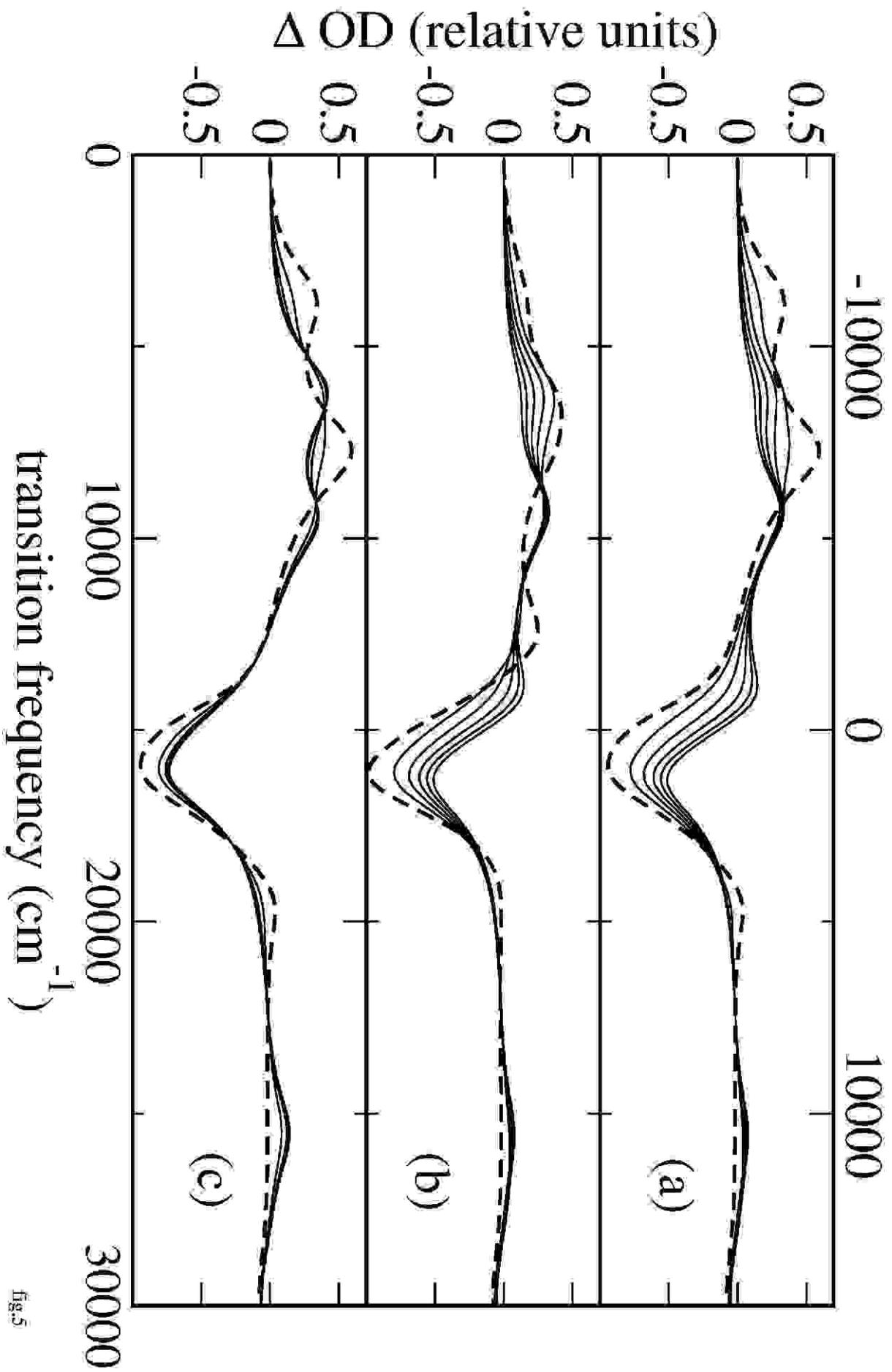

fig.5